# Framework for continuous transition to Agile Systems Engineering in the Automotive Industry

Jan Heine[1], Herbert Palm[2]

[1] Audi AG, Research & Development, Ingolstadt, Germany,
jan-hendrik.heine@audi.de

[2]University of Applied Sciences, Munich, Germany, herbert.palm@hm.edu

**Abstract:** The increasing pressure within VUCA-driven environments causes traditional, plan-driven Systems Engineering approaches to no longer suffice. Agility is then changing from a "nice-to-have" to a "must-have" capability for successful system developing organisations. The current state of the art, however, does not provide clear answers on how to map this need in terms of processes, methods, tools and competencies (PMTC) and how to successfully manage the transition within established industries. In this paper, we propose an agile Systems Engineering (SE) Framework for the automotive industry to meet the new agility demand. In addition to the methodological background, we present results of a pilot project in the chassis development department of a German automotive manufacturer and demonstrate the effectiveness of the newly proposed framework. By adopting the described agile SE Framework, companies can foster innovation and collaboration based on a learning, continuous improvement and self-reinforcing base.

## 1 Agile Systems Engineering – from "nice-to-have" to "must-have"

Traditional plan-driven, especially waterfall-like procedural models of systems development are no longer able to comprehensively cope with the pressure to innovate. New technologies at ever-shortening time intervals, the associated lack of technical experience, changing customer expectations and innovative business models - moving away from the traditional product towards services and the associated need for increasing integration of software with existing products and services - are creating an environment characterised by volatility, uncertainty, complexity and ambiguity (VUCA, [1]). The ability of a company to manage the challenges of this VUCA environment in its organisation and developmental setup is crucial for its future technical and economic success. Restricting agile development processes, methods, tools, and competencies (PMTC) to software developments alone is no longer sufficient. Agility is becoming key to successful Systems Engineering (SE) in all associated disciplines plus their interaction. The pressure for agile systems development is moving from being a "nice-to-have" to a "must-have" capability in many industry segments.

To date, state of the art does not provide a clear answer to this need. Since the first attempts to open up agile systems development [2], many approaches have emerged and (almost) as many have disappeared, with one of the biggest hurdles being the problem of scalability of agile Engineering from small SW to large system projects. The different approaches either originate in the world of agile software development (e.g. Scrum of Scrums [3], Large Scale Scrum [4], Scaled Agile Framework [5] etc.), in the plan-driven world (e.g. Water-Scrum-Fall [6]) or try to combine elements from both worlds from the outset. A uniform agile PMTC framework for the domain of system development is currently just as unrecognizable as a framework for its successful introduction.

Especially in the automotive industry, the accelerating pace of technological changes in form of powertrain electrification, highly automated driving, the use of artificial intelligence and the resulting increase in connectivity [7] causes a multitude of disruptive changes. OEMs and suppliers have recognised an unavoidable need for agile Systems Engineering within this VUCA environment by fundamentally adjusted PMTC approaches. Several companies prepared transformation programmes [8], whose success depend on both, the design of an adequate and agile PMTC framework and its successful introduction while simultaneously mastering day-to-day business.

In the following chapters, we present a concept for a new agile SE Framework for developments within the automotive industry including an inherent self-reinforcing component for successful introduction into ongoing development operations. In addition to other success factors, we highlight the need for project level gained learnings and knowledge to find their way from the operational level back to the process governance [9] in form of constructive feedback. This enables an iterative-incremental cross-project competence building as well as a continuous improvement and "agilisation" of the agile SE Framework per se. The organisation is able to learn, act proactively and respond to new challenges. In chapter 2, we define the elicited requirements for the agile SE Framework in the concrete development environment, before we explain our new methodological approach for realizing a continuous transition to agile SE in chapter 3. In Chapter 4, we describe the results and experiences of the piloting of our new approach, which we discuss in Chapter 5, and conclude with a brief outlook in Chapter 6.

## 2 Objectives of agile Systems Engineering

Agility is not an end in itself. The need to deliver system development project results in an agile way is driven by the need to define, design and implement systems effectively and efficiently within a VUCA environment. This vision drives a new methodological approach based on the ideas of agility and a learning organisation.

In addition to the need for a generic agile SE Framework that is aligned with VUCA requirements, process governance is also required for its continuation and sustainable use after its operationalisation. It is important to bear in mind that the introduction is taking place "at the open heart" with all risks of any big bang institutionalization. Therefore, not only the agile SE Framework as such is relevant, but also the process of its integration into the existing environment, which is predominantly determined by plan-driven actions. This results in the objectives listed in Table 1.

Table 1: Objectives

| Item | Objective: *The agile SE Framework…* |
|---|---|
| Self-reinforcement | …is intended to promote self-reinforcing transition to agile system development |
| SE at project level | …shall enable system development in the context of VUCA at project level (for both systems and subsystems) |
| V-Model compatibility | …shall allow system development considering the V-model |
| Cross-functional collaboration | …shall promote cross-functional collaboration (vs. working in silos) |
| Cross-project learning | …shall enable cross-project learning regarding product and process |
| Self-optimisation | …shall enable and promote feedback-based self-optimisation |
| Scalability | …shall allow scaling from one project to any number of projects |
| Guided introduction and maintenance | …shall provide for the necessary crucial roles for both its introduction and maintenance |
| Low-risk introduction | …must provide a concept for its low-risk introduction during ongoing day-to-day business as an integral component |

The stated objectives build the basis for designing the agile SE Framework.

## 3 New methodological approach for the agile SE Framework

Our newly designed agile SE Framework meeting the above-mentioned objectives comprises the four core elements indicated in Table 2.

Table 2: Four core elements (CE) of the agile SE framework.

| Core element (CE) | as characterized by… |
|---|---|
| CE1 – (agile) PMTC modular kit | consisting of (agile) PMTC elements and modules that are managed centrally at the level of process governance |
| CE2 – basic agile SE process | formalising the agile engineering process by using PMTC elements and producing feedback from project level to process governance |
| CE3 – Systems Scrum Master (role) | empowering the organisation and facilitating the project team |
| CE4 – self-reinforcing push-pull effect | enhancing the transition progress on macro-level through continuous top-down push (process governance) and bottom-up pull (projects) |

All four core elements - specifically tailored for systems development in the automotive industry but not restricted to it – will be described in detail below:

*The (agile) PMTC modular kit (CE1)* consists of a collection of agile elements or modules that form the ingredients for agile processes based on agile methods that can be executed by according techniques or tools for agile product development (including but not limited to systems definition, design and implementation) as well as project management and supporting tasks such as training and qualification. The elements are connectable and applicable in a modular manner to problems or projects. The PMTC modular kit is deployed from top-down in the organization through an institutionalized unit (process governance) that is responsible for its development, implementation and continuous improvement.

*The basic process for agile SE (CE2)* provides the basic agile problem problem-solving process representing elementary agile elements such as early feedback and continuous delivery. It serves as a basis for further project-specific tailoring. Fig. 1 shows this agile process flow inspired by the PDCA (Plan, Do, Check, Act) cycle [10], consisting of planning, (prototypical) execution, check, and adaptation steps, which are applied iteratively and incrementally in a Scrum-like manner [11] but also allow V-Model [12] integration.

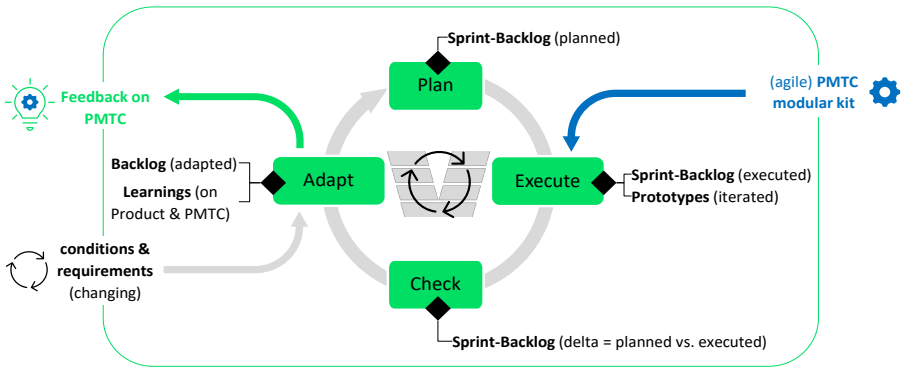

Fig. 1 basic process for agile SE

Individual cycles are initiated by a planning step ("Plan"), during which a Sprint-Backlog is defined. In the following execution step ("Execute"), a prototypical realization of the planned Sprint-Backlog is carried out and tested in a protected environment in order to gain learnings regarding product and applied PMTC modular kit elements. Based on these learnings, along with potentially changed conditions and requirements, measures for the intended realization in the live environment are identified during an adaption step ("Adapt"), resulting in an adapted Backlog. Finally, product-related decisions are taken and feedback regarding PMTC is provided to the process governance to optimize the agile SE Framework.

The overall progress can be measured ("Check") both across sprints using the different configurations of the adapted Backlog over time and "on the fly" using the Backlog progress.

*The critical role of the Systems Scrum Master (CE3)* reinterprets the classic roles of the Systems Engineer [13] and the Scrum Master [14] in the context of agile systems development. Their mission is to empower the organisation to get started and to actively maintain and support the basic process for agile SE within a project. The Systems Scrum Master both enforces and enables teams to make use of the (agile) PMTC modular kit. This includes qualifying teams, organizing interdisciplinary engineering activities and synchronizing teams. The Systems Scrum Master is responsible for ensuring adherence to the basic process and facilitate the bottom-up learning feedback flow to the governance unit.

*The self-reinforcing push-pull effect (CE4)* enables a systematic step-by-step transition from a plan-driven organization and mode of operation to agile SE over time. Its basic principle is illustrated in Fig. 2. Initially, new (agile) PMTC modular kit's elements are pushed top-down through the organization into individually selected projects. To avoid the risks of a big bang "flooding", it is necessary to introduce the transition to agile SE within a few selected environments (i.e. within pilot projects) and make use of their success stories. These success stories create a pull character to new projects. At the same time, the PMTC modular kit is being revised by the process governance unit based on constructive bottom-up feedback from project level, making it more suitable – and agile – for projects with increasing bandwidth.

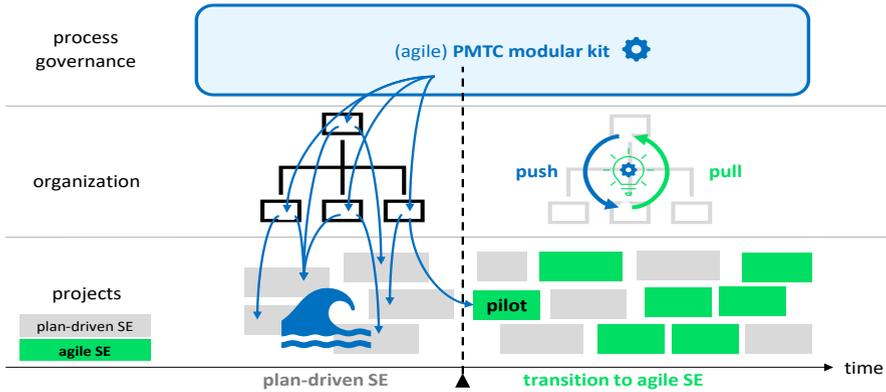

Fig. 2 Cross-organizational self-reinforcing push-pull effect

The growing maturity of the agile SE framework means that other projects can be involved in the new approach, creating a scaling effect and providing additional feedback. Through the continuous interaction of top-down push from the process governance and bottom-up pull from the operational projects, the agile SE Framework can continuously spread and provide increasingly suitable solutions as well as increasing cross-organizational knowledge about their applicability in real-life project environment. This ultimately leads to a self-reinforcing effect and thus continuous improvement of projects' effectiveness and efficiency.

## 4 Piloting the agile SE Framework – Proof of Concept

In the following, we describe the introduction within a pilot project, its results and the emerging push-pull effect of the agile transformation.

The piloting took place in the chassis development area of a German automotive manufacturer. A novel and safety-critical "by-wire" system was addressed, involving multiple engineering disciplines. A ten-member project team from the OEM and development service providers contributed technical and methodological expertise, albeit with limited availability due to concurrent tasks. Fig. 3 sketches the four core elements of the agile SE Framework being implemented and adapted over 16 two-week development increments. The initial increment was carried out in a more plan-driven way, while the subsequent increments became increasingly agile. By combining iterative learning with individual training units, the team experienced a steep learning curve enabling it to produce a large number of product and PMTC-related results. Due to time constraints and the overall project's early stage, the product-related deliverables focused on those of the design branch ("left side") of the V-model. The PMTC-related findings were fed back to the process governance in a constructive way as basis for further optimisation. Identified success factors were communicated within the organisation and its management in form of project's success.

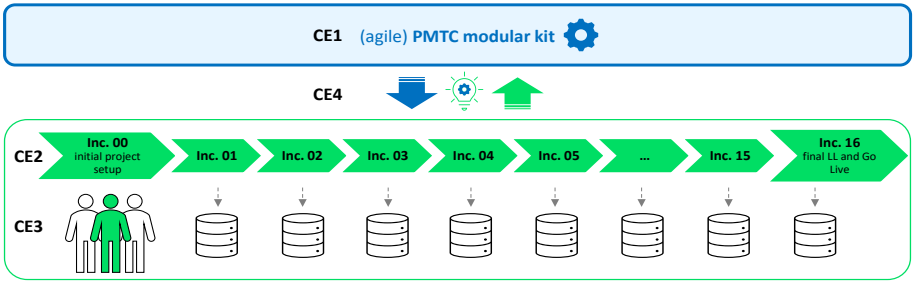

Fig. 3 Phase sequence of the pilot project

The tangible benefits in terms of quality (especially transparency as a base for consistency) of the engineering artefacts, increased efficiency, strong process capability in reference to Automotive SPICE, competence building, and a developing team spirit convinced another three project teams to follow up. The pilot project, therefore, successfully underlined both the capability of the agile SE Framework concept with its defined agile PMTC elements and its agile transition approach based on the self-reinforcing push-pull effect.

# 5 Key success factors

During the proof of concept phase, we identified key success factors as well as possible obstacles of the newly defined agile SE Framework related to its successful implementation:

*Qualification and competence building:* Ideally, the team is already trained in the framework basics and continually strengthens its competences. Thereby, cultivating systems thinking, as opposed to isolated expertise, plays a central role. In addition, the team benefits from correspondence with other teams qua ongoing PMTC modular kit improvements.

*Appropriate process model:* The process model allows early feedback-based learning on the product and PMTC elements. Iterative optimisation of the project-specific setup gradually increases the maturity of both the product and the used PMTC elements. In addition, the measurability and visualisation of the progress of results builds confidence within the project team and trust with external stakeholders.

*Step-by-step introduction:* A step-by-step introduction enables the incremental and value-driven replacement of existing plan-based ways of working with agile surrogates, taking into account the given actual development environment. Early identification and mitigation of emerging risks reduces the overall risk of a "big bang" rollout to an acceptable minimum.

*Guidance and facilitation:* Guidance and support from the beginning enables the team to overcome initial hurdles and possible scepticism. The team can learn and adapt to the new agile methods faster and more effectively and the organisation is enabled to put theory into practice. Furthermore, the new way of interdisciplinary collaboration requires a facilitator, represented by the Systems Scrum Master(s). Their ability to maintain a balanced triad of technical, methodological and social skills is crucial.

*Create a self-reinforcing momentum:* It is of crucial importance to communicate benefits and value of the agile framework experiences gained in projects ("talk the walk"). At the same time, these stories have to trigger new projects and applications ("walk the talk"). Value is persuasive and leads to trust, enthusiasm and the "necessary mindset" among those involved, creating a self-reinforcing momentum for successful transformation.

## 6 Concluding remarks

The proposed agile SE Framework represents a pioneering approach to the adoption of agility on system development level in the automotive industry. A major focus of this paper is to substantiate the implementation of a newly defined agile SE Framework. By implementing this framework, system companies can foster innovation, collaboration and continuous improvement to adapt and thrive in an ever-changing VUCA driven landscape. Four core elements are forming the framework's backbone: An agile PMTC modular kit, a basic process for agile SE, the role of the System Scrum Master and a self-reinforcing transition concept. Results of a first pilot are reported together with identified success factors. They lay the foundation for the scaling and application of the framework in other projects and organisations. Many questions are still open. Further experience must be gathered and evaluated. But a promising start has been made.